\documentclass[aps,prd,showpacs]{revtex4}
\usepackage{graphicx}
\usepackage{revsymb}
\begin{document}
\renewcommand{\thesection}{\arabic{section}}
\renewcommand{\thesubsection}{\arabic{subsection}}
\title{ Saha Equation for the
Photo-Ionization of Hydrogen Atoms in Partially Ionized Relativistic Hydrogen Plasma and the Effect of Gravity 
on the Binding of Hydrogen Atoms in Rindler Space}
\author{Sanchita Das$^{a,1)}$ and Somenath Chakrabarty$^{a,2)}$}
\affiliation{
$^{a)}$Department of Physics, Visva-Bharati, Santiniketan 731 235, 
India\\ 
$^{1)}$Email:sanchitadas.rs@visva-bharati.ac.in\\
$^{2)}$Email:somenath.chakrabarty@visva-bharati.ac.in
}
\date{\today}
\begin{abstract}
We have studied Saha equation for photo-ionization of hydrogen atoms in 
partially ionized relativistic hydrogen plasma in Rindler space. Following the principle of equivalence, we have 
obtained the abundances of neutral hydrogen atoms, hydrogen ions and the electrons in dynamic equilibrium of the 
photo-ionization reaction of neutral hydrogen atoms and electron capture process by hydrogen ions (de-ionization
process) and also investigated their 
variations with temperature of the plasma and the uniform gravitational field in the Rindler space or equivalently
the uniform acceleration of the observer. Hence obtained the Saha ionization formula for partially ionized hydrogen
plasma in Rindler space. It has been observed that the abundance of neutral hydrogen atoms
decreases with the increase in temperature of the plasma, which is the usual picture, 
whereas it increases with the increases in the strength of
uniform gravitational field. The second part of this observation shows that the binding of
the electrons inside hydrogen atoms increases
with the increase in the strength of gravitational field or equivalently an observer with
very high acceleration will see less amount of ionized hydrogen atoms compared to inertial
observer.
\end{abstract}
\maketitle
%\noindent Keywords:
%Saha equation, Uniformly accelerated frame, Inertial frame, Rindler space, Rindler coordinates, Photo ionization, Quantum binding
\section{Introduction}
It is  well known that the conventional Lorentz transformations are
the space-time coordinate transformations between two inertial
frame of references 
\cite{LL}.
However, following the principle of equivalence, it is trivial to obtain the
space-time transformations between a uniformly accelerated frame 
and an inertial frame and vice-versa
in the same manner as it is done in special theory
of relativity
\cite{WB,MTW,RL,MO,BD}. In the present scenario the flat space-time geometry is
called the Rindler space.  
For the sake of  illustration of principle of equivalence,
one may state, that a
reference frame undergoing an accelerated motion in absence of gravitational
field is equivalent to a frame at rest in presence of a 
gravitational field. Therefore in the present picture, the magnitude of the uniform
acceleration is exactly equal to the strength of gravitational field. 
We may assume that the gravitational field is produced by a strong gravitating
object. We further approximate that the gravitational field is
constant within a small domain of spacial region. Since it is exactly equal to the inform
acceleration of the moving frame, this is also called the local acceleration
of the frame.

To study Saha equation in an uniformly accelerated frame of
reference or in Rindler space,
we first develop a formalism 
with the physical concepts of principle of equivalence as discussed above,
and 
obtain the elements of the metric tensor
$g^{\mu\nu}$. 
We shall show that
analogous to the Minkowski space-time metric tensor, the off-diagonal 
elements of $g^{\mu\nu}$ are also zero in Rindler space.
In the next step, with the 
conventional form of action as has been defined in special
theory of relativity \cite{LL}, which is invariant here also,
the Lagrangian of a particle (in
our study, which may 
be a hydrogen atom or a hydrogen ion or an electron) is derived from
Hamilton's principle. Which
further gives the momentum and energy or Hamiltonian of the particle from the
standard relations of classical mechanics. Then considering a partially 
ionized 
hydrogen plasma which is a reactive mixture of neutral hydrogen atoms, 
hydrogen ions, electrons and
photons in dynamic chemical equilibrium, we shall obtain the modified
form of Saha equations when observed from  a uniformly 
accelerated
frame of reference or in Rindler space. 
To the best of our knowledge, the study of relativistic version of Saha
equation in Rindler space  has not been reported earlier.
We shall also compare our findings with the conventional results.

We have organized the article in the following manner: In the next section, for the sake of
completeness, 
we shall give a brief review from the existing literature 
to obtain some of the useful relations of special theory of
relativity in uniformly accelerated frame. In this section we shall
also obtain the relativistic version of Hamiltonian in Rindler
coordinate system.
In section-3 we shall investigate photo-ionization of hydrogen atoms
in partially ionized hydrogen plasma in Rindler space. The expression for the number
densities are obtained in Appendix-A. To obtain the number densities, we assume for the sake
of simplicity that the constituents of the reactive mixture, which is a partially ionized
relativistic hydrogen plasma in Rindler space, 
behave classically, i.e., they obey  the relativistic version of
Boltzmann distribution. 
In section-4 we  have obtained the relativistic version of Saha ionization formula in
Rindler space. We have shown that the concentration of neutral hydrogen atoms decreases with
the increase in temperature, which is the conventional scenario, whereas it increases with the
increase in the strength of gravitational field or the magnitude of acceleration of the
non-inertial frame. The gravitational field, which is a classical entity enters in the
quantum problem through the Rindler Hamiltonian.
In the last section we give the conclusion of our findings.
%%%000000000000000000000000000000000000000000000000000000
\section{Basic Formalism}
In this section, for the sake of completeness, 
following the references \cite{MS,MAX,DP} we shall establish some of the useful formulas of
special theory of relativity for a uniformly accelerated frame of
reference. Before we go to the scenario of uniform acceleration of the moving frame, 
let us first assume that the frame $S^\prime$  has rectilinear motion with
uniform velocity $v$ along
$x$-direction with respect to some inertial frame $S$. Further the 
coordinates
of an event occurred at the point $P$ (say) is indicated by
$(x,y,z,t)$ in $S$-frame and with
$(x^\prime,y^\prime,z^\prime,t^\prime)$ in the frame $S^\prime$. The
primed and the un-primed coordinates are related by the conventional form
of Lorentz transformations and are given by
\begin{eqnarray}
x^\prime&=&\gamma(x-vt), ~~y^\prime=y,~~ z^\prime=z ~{\rm{and}}~ \nonumber \\
t^\prime&=&\gamma\left( t-vx\right) ~{\rm{with}}~
\gamma=\left (1-v^2\right )^{-1/2}
\end{eqnarray}
is the well known Lorentz factor. Throughout this article we have followed the natural system of 
units, i.e., speed of light in
vacuum, $c=1$ and later we put the Boltzmann constant $k_B=1$ and the Planck constant $h=1$. 
Next we consider a uniformly accelerated
frame $S^\prime$ moving with uniform acceleration $\alpha$ also along $x$-direction in $S$-frame. 
Then the Rindler coordinates are given by (see the references \cite{MS,MAX,DP}),
\begin{eqnarray}
t&=&\left (\frac{1}{\alpha}+x^\prime\right )\sinh\left (\alpha t^\prime
\right ) \nonumber  ~~{\rm{and}}~~ \\
x&=&\left (\frac{1}{\alpha}+x^\prime\right )\cosh\left (\alpha t^\prime
\right ) 
\end{eqnarray}
Hence one can also express the inverse relations
\begin{equation}
t^\prime=\frac{1}{2\alpha}\ln\left (\frac{x+t}{x-t}\right )
~~{\rm{and}}~~ x^\prime=(x^2-t^2)^{1/2}-\frac{1}{\alpha}
\end{equation}
The Rindler space-time coordinates as mentioned above
are then just an accelerated frame
transformation of the Minkowski metric of special relativity. The
Rindler coordinate transformations change the Minkowski line element from
\begin{eqnarray}
ds^2&=&dt^2-dx^2-dy^2-dz^2   ~~{\rm{to}}~~ \\ ds^2&=&\left
(1+\alpha x^\prime\right)^2{dt^\prime}^2-{dx^\prime}^2
-{dy^\prime}^2-{dz^\prime}^2
\end{eqnarray}
Since the motion is assumed to be rectilinear and along $x$-direction, 
$dy^\prime=dy$ and $dz^\prime=dz$. The form of the
metric tensor can then be written as
\begin{equation}
g^{\mu\nu}={\rm{diag}}\left (\left (1+\alpha x\right
)^2,-1,-1,-1\right )
\end{equation}
Since we shall deal with the accelerated frame only, we have dropped the prime symbols.
Now following the concept of kinematics of particle motion in special theory
of relativity \cite{LL}, the action
integral may be written as (see also \cite{CGH} and \cite{DLM})
\begin{equation}
S=-\alpha_0 \int_a^b ds\equiv \int_a^b Ldt
\end{equation}
Then using eqns.(5) and (7) and putting $\alpha_0=-m_0$ \cite{LL}, where $m_0$ 
is the
rest mass of the particle, the Lagrangian of the particle is given by
\cite{DLM}
\begin{equation}
L=-m_0\left [\left ( 1+\alpha x\right )^2 -v^2
\right ]
\end{equation}
where $v$ is the velocity of the particle. The momentum of the
particle is then given by
\begin{equation}
p=m_0 v\left [ \left (1+\alpha x \right )^2
-v^2 \right ]^{-1/2}
\end{equation}
Hence the Hamiltonian of the particle or the single particle energy is 
given by
\begin{equation}
H=\varepsilon(p)=m_0 \left (1+\alpha x \right ) \left (1+
\frac{p^2}{m_0^2}\right )^{1/2}
\end{equation}
This is the well known Rindler Hamiltonian.
In the next sections, using this simple form of
single particle energy, we shall obtain the number densities for the constituents of 
partially ionized
relativistic hydrogen plasma.
\section{Relativistic Kinetic Theory for Partially Ionized Hydrogen Plasma in Rindler Space}
We assume that the constituents of the plasma are neutral hydrogen, hydrogen ions and electrons. This is a reactive
mixture along with photons, which are colliding with neutral hydrogen atoms and excite or ionize 
them. All of these constituents are in
thermodynamic equilibrium in Rindler space. The single particle energy for all the constituents, except the photons
are  given by eqn.(10). In this expression only the rest mass will
be different for different constituents. Then following \cite{de} the expression for number density may be written
as 
\begin{equation}
n(T,u)=\frac{gz}{(2\pi)^3} \int d^3p \exp[-\beta u(\alpha)(p^2+m_0^2)^{1/2}]
\end{equation}
where $u(\alpha)=1+\alpha x$ and it is further assumed that within the domain denoted by $x$, the gravitational
field $\alpha$ is constant. Moreover, to avoid the arbitrariness of $\alpha$ and $x$, we
assume $u$ as the independent variable.
Here $\beta=1/T$, $g$ is the degeneracy of the constituent and $z=\exp(\mu/T)$ is its fugacity. The above integral over momentum
can be evaluated in terms of modified Bessel function of second kind of order two. In Appendix-A we have given a
brief outline of its evaluation. Then we have 
\begin{equation}
n(T,u)=\frac{gz}{2\pi^2} Tm_0^2\frac{K_2\left (\frac{m_0 u}{T}\right )}{u}
\end{equation}
In the above expression, if we put $u=1$, i.e., $\alpha=0$, we get back the conventional form of the expression for
number density.
Hence one can very easily show that the chemical potential may be expressed as
\begin{equation}
\mu=T\ln \left [\frac{n(T,u)2\pi^2u}{gTm_0^2}\frac{1}{K_2\left ( \frac{m_0u}{T}\right )}\right ]
\end{equation}
\section{Relativistic Version of Saha Ionization Formula in Rindler Space}
In this section we shall study the photo disintegration of hydrogen atoms in a partially
ionized relativistic hydrogen plasma in Rindler space. We assume that the constituents
(neutral hydrogen, hydrogen ions and electrons) are in thermodynamic equilibrium. We further
assume that because of high temperature of the plasma, electrons inside the neutral hydrogen
atoms are not necessarily in the ground state. Let us assume that they are in some $n$th
bound excited state ($n>1$) and the corresponding degeneracy is $g_n$ (see \cite{PH} and
\cite{RQKN} for the non-relativistic calculation. Whereas in \cite{DE} a preliminary studies has been done for the
Saha ionization in the Rindler space. See also \cite{SKRA} for a related interesting work).
We start with the ionization (de-ionization or capture process), given by
\begin{equation}
H_n+\gamma \leftrightarrow H^++e^-
\end{equation}
where the index $n$ indicates that the neutral hydrogen is in the $n$th bound excited state.
We further assume that the constituents are in dynamic chemical 
equilibrium. In this condition the rates for the ionization process and the de-ionization process are exactly equal. 
Of course the equilibrium point is a function of
both temperature of the plasma and the strength of uniform gravitational field. The condition for dynamic
chemical equilibrium is given by
\begin{equation}
\mu_{H_n}=\mu_{H^+}+\mu_e
\end{equation}
Because of the non-conservation of photon numbers in the reactive mixture, the chemical potential of photons do
not appear in the chemical equilibrium condition, i.e., $\mu_gamma=0$.
Using the expression for chemical potential as given by eqn.(16) and using the chemical equilibrium condition, given
by eqn.(14) and further assuming that the rest mass of $H_n$ atom
and hydrogen ion are equal, we can write
\begin{equation}
R(T,u)\approx \frac{n_{H_n}}{n_{H^+} n_e}=C \frac{u}{T}\frac{1}{K_2\left ( \frac {m_e u}{T}\right )}
\end{equation}
where $m_e$ is the rest mass of electron and $C$ is a constant, a function of the degeneracies of $H_n$ atom,
hydrogen ion and electron. This is the relativistic version of Saha ionization formula. This gives the equilibrium
ratio of number density of $H_n$ atoms and the product of the number densities of hydrogen ions and the electrons.  
With $u=1$ or $\alpha=0$ we get back the conventional result for Saha ionization in the relativistic scenario in
Minkowski space. In fig.(1) we have shown the variation of $R(T,u)$ with temperature in MeV for $u=1$ (the solid
curve) and with $u$ for
$T=5$MeV (dashed curve). The solid curve of this figure shows that the ratio increases with 
temperature, which is the conventional
scenario. Or in other ward, we can say that the number density of neutral hydrogen atoms
decreases with the increase in temperature. It actually indicates that 
the thermal ionization rate will increase with the increase in temperature of the plasma.
The dashed curve indicates the variation of the ratio with $u$. It indicates that the
electrons inside the neutral atoms become more bound in presence of strong gravitational
field.
The ionization of hydrogen atom in the ground state or in any one of the $n$th. bound excited
state does not in the expression for the ratio of number densities because of the
relativistic nature of the constituents. The kinetic energies are quite high compared to the
ionization energy. The later is a few eV. Of course in the non-relativistic approximation it
will appear in the expression for the ratio $R(T,u)$.

Lastly we would like to make non-relativistic approximation when $r_0$ is large or
temperature is low enough. We shall now show that the ionization potential will appear
automatically in Saha ionization formula. Using the
dynamical form of chemical equilibrium, given by eqn.(14) and without considering any approximation made
previously, we have 
\begin{equation}
\frac{n_{H_n}}{n_{H^+}n_e}\left (\frac{m_{H^+}m_e}{m_{H_n}}\right )^2\frac{g_{H^+}g_e}{g_{H_n}}
T\frac{K_2\left (\frac{m_{H^+}}{T}\right )K_2\left (\frac{m_e}{T}\right)}{K_2\left (\frac{m_{H_n}}{T} \right )}=1
\end{equation}
Now for large values of the arguments of the modified Bessel functions, we have
\begin{equation}
K_2(r_0)\approx\left (\frac{\pi}{2}\right )^{1/2} r_0^{-1/2}\exp(-r_0)
\end{equation}
(see eqn.(46) in page 49 of reference \cite{de}). Hence we have
\begin{equation}
\left [\frac{n_{H_n}}{n_{H^+}n_e}\left (\frac{m_{H^+}m_e}{m_{H_n}}\right )^{1/2}\frac{g_{H^+}g_e}{g_{H_n}}
\frac{T^{3/2}}{(2\pi u)^{3/2}} \right ]\exp(-\varepsilon_n/T)=1
\end{equation}
where $\varepsilon_n=m_{H^+}+m_e-m_{H_n}$, the ionization potential for $H_n$ atom. Hence we can write
\begin{equation}
R(T,u)=\frac{n_{H_n}}{n_{H^+}n_e}\propto \frac{u^{3/2}}{T^{3/2}}\exp(-\varepsilon_n/T)
\end{equation}
Which is the expression in the non-relativistic approximation. If we assume that the
temperature is low enough and is constant, then the ratio $R(T,u)\propto u^{3/2}$. Which
also indicates that with the increase of $u$, the density of neutral hydrogen atoms
increases, or in other wards, the binding of electrons inside hydrogen atom increases with
the gravitational field in the Rindler space.
\section{Conclusion}
Since most of the vital points of this work has already been discussed in the main text,
here we shall give a very brief conclusion of our work. We have studied Saha ionization in a
partially ionized relativistic hydrogen plasma in Rindler space. It has been observed that
the conventional results are fooled automatically if the presence of gravitational field is
switched off. Further we have noticed that the electrons in the hydrogen atoms become more
strongly bound with the increase in the strength of gravitational field in the Rindler
space. As a future study we shall report the energy eigenvalue problem of hydrogen atom in
Rindler space and show that the binding energy of electrons inside hydrogen atom increases
linearly with the strength of gravitational field in Rindler space (see \cite{MITRA} for some preliminary
calculation,  see also \cite{BRAU}).
\section{Appendix-A}
The expression for the number density is given by
\begin{equation}
n(T,u)=\frac{g}{(2\pi)^3} \int d^3p z\exp[-\beta u(\alpha)(p^2+m_0^2)^{1/2}]
\end{equation}
Let us substitute
\[
\frac{u(\alpha)(p^2+m_0^2)^{1/2}}{T}=r
\]
Then
\[ p=\frac{T}{u}(r^2-r_0^2)^{1/2}
\]
where $r_0=m_0 u/T$. Hence
\[
dp=\frac{T}{u}\frac{r}{(r^2-r_0^2)^{1/2}} dr
\]
Substituting in the expression for number density, we have
\[
n(T,u)=\frac{gz}{2\pi^2}\frac{T^3}{u^3}\int_{r_0}^\infty  exp(-r)(r^2-r_0^2)^{1/2} r dr
\]
Finally using the integral representation of $K_\nu(r_0)$, given by
\[
K_\nu(z)=\frac{2^{\nu-1}(\nu-1)!}{(2\nu-2)!}\frac{1}{r_0^\nu}\int_{r_0}^\infty dr (r^2
-r_0^2)^{\nu-3/2} r \exp(-r)
\]
we have (see eqn.(32), page 48 of reference \cite{de})
\begin{equation}
n(T,u)=\frac{gz}{2\pi^2} \frac{Tm_0^2}{u} K_2\left (\frac{m_0u}{T} \right )
\end{equation}
In the derivation of Saha ionization formula we shall use this expression for number density for the various
constituents with different $m_0$.
Substituting the value of fugacity $z$, we can express the chemical potential in terms of
the number density, temperature $T$ of the plasma and the strength of gravitational field $u$ and is given by
\begin{equation}
\mu=T[\ln(2\pi^2)-\ln(T)-\ln(g)-2\ln(m_0)+\ln(n)-\ln K_2\left (\frac{um_0}{T}\right )+\ln(u)
]
\end{equation}

%%%%%%%%%%%%%%%%%%%%%%%%%%%%%%%%%%%%%%%%%%%%%%%%%%
\begin{figure}[ht]
\centerline{\includegraphics[width=4.0in]{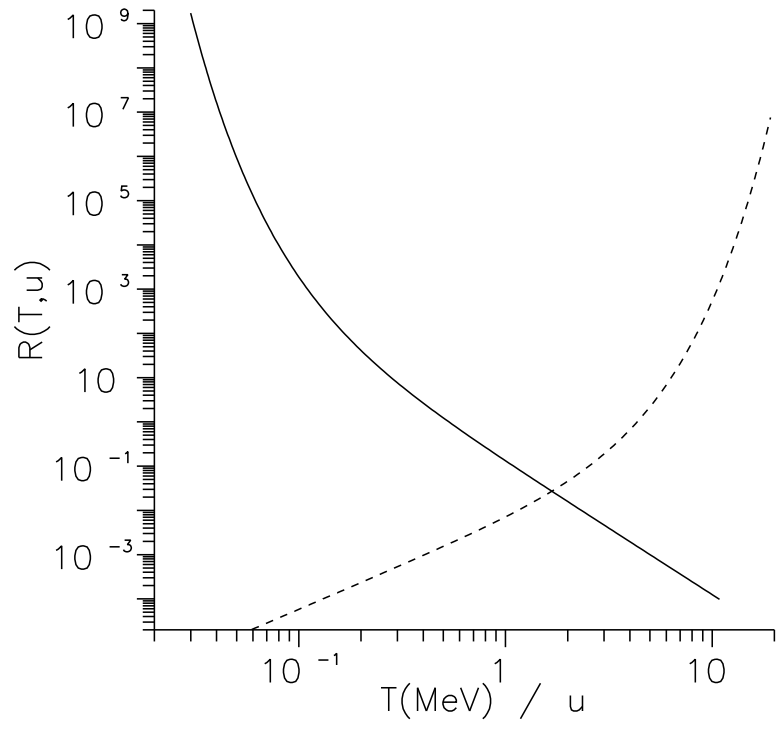}}
\caption{Variation of the ratio $R(T,u)$ with $T$ (solid curve) and $u$ (dashed curve)}
\end{figure}
%%%%%%%%%%%%%%%%%%%%%%%%%%%%%%%%%%%%%%%%%%%%%%%%%%%%%%%%%

\begin{thebibliography}{}
\bibitem{LL} L.D. Landau  and E.M. Lifshitz,
The Classical Theory of Fields, Butterworth-Heimenann, Oxford, (1975.)
\bibitem{WB} S. Weinberg, Gravitation and
Cosmology,Wiley, New York, (1972).
\bibitem{MTW} C.W. Misner,  Kip S. Thorne  and
J.A. Wheeler, Gravitation, W.H. freeman and Company, New York, (1972).
\bibitem{RL} W. Rindler, Essential Relativity,
Springer-Verlag, New York, (1977).
\bibitem{MO} C. Moller, The Theory of Relativity,
Calarendon Press, Oxford, (1972).
\bibitem{BD} N.D. Birrell and P.C.W. Davies,
Quantum Field Theory in Curved Space, Cambridge University Press,
Cambridge, (1982).
\bibitem{MS} M. Socolovsky, Annales de la
Foundation Louis de Broglie {\bf{39}}, 1, (2014).
\bibitem{MAX} G.F. Torres del Castillo Torres  and C.L. Perez
Sanchez, Revista Mexican De Fisika, {\bf 52}, 70 (2006).
\bibitem{DP} D. Percoco and V.N. Villaba,  Class. Quantum Grav.,
{\bf{9}}, 307, (1992).
\bibitem{CGH} C-G Huang and J-R Sun,
arXiv:gr-qc/0701078,(2007).
\bibitem{DLM} Domingo J. Louis-Martinez,
Class Quantum Grav., {\bf{28}}, 035004, (2011). 
\bibitem{de} Relativistic Kinetic Theory: Principles and Applications, S.R. de Groot, W.A. van Leeuwen and 
Ch.G. van Weert, North Holland Pub. Co., Amsterdam and New York, 1980. 
\bibitem{PH} A.C. Phillips, The Physics of Stars, John
Wiley \& Sons, New York, (1996).
\bibitem{RQKN} Stellar Astrophysics, R.Q. Huang and K.N. Yu, Springer, 1998.
\bibitem{DE} Sanchari De and Somenath Chakrabarty, Pramana Jour. Phys., (2017) 88:89. DOI
10.1007/s12043-017-1421-0
\bibitem{SKRA} S. Kichenassamy and R.A. Krikorian, J. Phys. A: Math. Gen. {\bf{16}}, 2347 (1983).
\bibitem{MITRA} Soma Mitra and Somenath Chakrabarty, EPJ Plus, 71, 5, 2017.
\bibitem{BRAU} F. Brau, arXiv:hep-ph/9903269; JMP {\bf{40}}, 1119, (1999).
\end{thebibliography}
\end{document}